\documentclass[11pt]{article}
\usepackage{latexsym}
\usepackage{epsfig,amssymb,euscript}
\usepackage{amsmath}
\usepackage{array,calc,epsfig}
\usepackage{graphicx}
\usepackage{dcolumn}
\usepackage{bm}

\def\be{\begin{equation}}
\def\ee{\end{equation}}
\def\bea{\begin{eqnarray}}
\def\eea{\end{eqnarray}}

\begin{document}
\leftline{\tt hep-th/0703284}

\vskip -.8cm

\rightline{\small{\tt UB-ECM-PF 07/06}}
\vspace{20pt}

\begin{center}
{\Large{\bf  Splitting of macroscopic fundamental strings in flat
space and holographic hadron decays}}
\vskip 0.7 cm
\vspace{10pt}

{\Large{F. Bigazzi}}\\
\vspace{12pt}
{\it Physique Th\'eorique et Math\'ematique and International Solvay
Institutes, Universit\'e Libre de Bruxelles; CP 231, B-1050
Bruxelles,
Belgium. \\ fbigazzi@ulb.ac.be}\\
\vspace{16pt}
{\Large{A. L. Cotrone}}\\
\vspace{12pt}
{\it Departament ECM, Facultat de F\'isica, Universitat de Barcelona and Institut
de Fisica d'Altes Energies, Diagonal 647, E-08028 Barcelona,
Spain. \\ cotrone@ecm.ub.es}\\
\vspace{16pt}
{\Large{L. Martucci and W.Troost}} \\
\vspace{12pt}
{\it Institute for theoretical physics, K.U. Leuven,
Celestijnenlaan 200D, B-3001 Leuven,
Belgium. \\ luca.martucci@fys.kuleuven.be,
walter.troost@fys.kuleuven.be}
\end{center}

\vspace{12pt}

\begin{center}
\textbf{Abstract}
\end{center}

\vspace{4pt} {\small \noindent
In this review article we present the calculation of the splitting
rate in flat space of a macroscopic fundamental string either
intersecting at a generic angle a Dp-brane or lying on it. The
result is then applied, in the context of the string/gauge theory
correspondence, to the study of exclusive decay rates of large
spin mesons into mesons. As examples, we discuss the cases of
${\cal N}=4$ SYM with a small number of flavors, and of QCD-like
theories in the quenched approximation. In the latter context,
explicit analytic formulas are given for decay rates of mesons
formed either by heavy quarks or by massless quarks.}

\section{Introduction}
Since the mid--seventies, when it was discovered
\cite{Scherk:1974ca} that quantized string theory incorporates
gravity, this fact has been the main agent pushing forward the
development of the theory, even if quantum gravity has not yet
come within the realm of tomorrows experiments. Whereas some very
surprising structural properties were uncovered, these have not
(yet) brought string theory to the level of presently testable
predictions as applied to gravity or the universe. The discovery
of, most notably,
 branes \cite{Dai:1989ua},  and the string/gauge theory correspondence \cite{Maldacena:1997re}
have however also revitalized the more ancient arena for the
theory of strings, namely strong interaction physics. Under the
accepted paradigm of quantum chromodynamics with color
confinement, the latter is realized  by the formation of a color
flux tube. String theory provides a model for such color flux
tubes. In this way, developments in `pure' string theory alluded
to above can be put to use for the `applied' string theory of
hadronic physics. At present, no single standard model of strings
(and branes) for strongly interacting particles has been agreed
upon: string/brane models do not exactly correspond to the
physical world, and parameters in manageable computations are not
always in the measured physical ranges, to put it mildly.
Nevertheless we think that hadronic physics is a natural arena to
put the tricks and tools of string technology to use.

A variety of hadronic properties have been approached from this
perspective. Using the string/gauge theory correspondences,
various brane settings have been proposed that mimic at least
partly a $SU(N)$ color gauge theory with quarks. Spectra (see for example \cite{myers1})
and scattering processes, both inclusive \cite{ps2} and exclusive \cite{ps1}
were investigated. On a perhaps more model independent level, the
question of the color flux tube model itself (thin strings vs.
fat strings) received attention in \cite{susskind}.

In this paper, we review our treatment \cite{noi1,noi2} of meson decays, where
mesons are pictured as a quark and an antiquark at opposite ends
of a color tube. The way that quarks and flux tubes are
represented differs, depending on the concrete incorporation of
QCD (or a QCD-like theory) into a brane model. Always the decay of the meson (into other
mesons) is governed by the splitting of the string that represents
the color tube. The splitting happens as a consequence of the
string crossing another brane, that is required to incorporate
different flavors into the model. Therefore, in section 2, we start by discussing  the general question of splitting
rates when it intersects a brane, following the pattern set in
 \cite{p,dp}. In section 3, after the general treatment is
introduced, a first application is made in which
mesons are represented as open strings attached to D7-branes, placed in the $AdS_{5}\times S^{5}$ gravity 
background sourced by $N\gg 1$
D3-branes, at positions that are correlated with the quark
masses of the different flavors. The calculation (unfortunately)
requires an approximation where the number of flavors is much
smaller than the number of colors. It applies to the decay of
high spin mesons $\bar{Q} Q$  into a pair of mesons $\bar{Q}q$ and
$\bar{q}Q$, where the quarks $q$ and $\bar{q}$ created in the decay
have mass $m_q<<m_Q$. In a second model, a stack of ($N\gg 1$)
$D4-$branes provides the gauge theory background to which $N_f$
$D6-$ \cite{myers} or $D8-$ \cite{ss} branes add flavor. For both cases,
we model the decay $\bar{Q}Q\rightarrow \bar{Q}q + \bar{q}Q$ and
$\bar{q}q\rightarrow \bar{q}q + \bar{q}q$ respectively, where the
latter involves light quarks only.

\section{Splitting of macroscopic strings in flat space}

In this section we provide the formulas for the splitting rate of a macroscopic open string either intersecting at generic angle $\theta$ a generic Dp-brane, or lying on a generic Dp-brane, in flat space \cite{noi1,noi2}.
The two computations are very similar and can be carried out at the same time.

\begin{figure}[th]
\centerline{\psfig{file=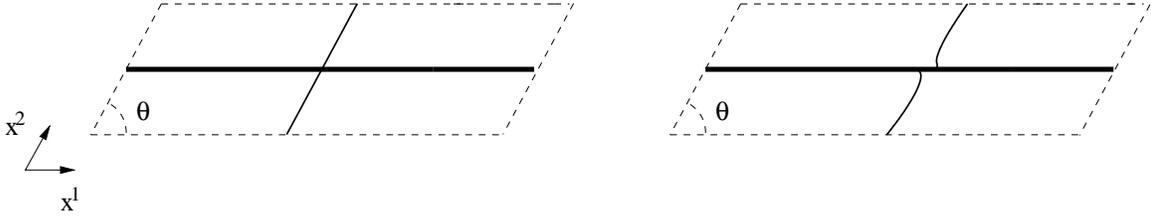,width=6.0in}}
\vspace*{8pt}
\caption{The two torus $T^2_\theta$. The thick line is the brane, the slim line is the string. On the right, the string 
after the splitting.\protect\label{torus}}
\end{figure}

A very useful trick to set up this computation is, following
 \cite{p,dp,jjp}, to use the S-matrix formalism and an optical
theorem, deriving thus the rate from the forward amplitude for the
propagation of the string. The latter can be obtained from a
vertex operator matrix element after compactifying some space
dimensions on a (very large) torus.  In the case where the string
intersects the Dp-brane, we compactify on $R_t \times
T^2_\theta\times T^{p-1}_{\|}\times T^{8-p}_{\perp}$, where $R_t$
refers to the time direction. The  $T^2_\theta$ is parameterized
by $(x^1,x^2)\simeq (x^1+n_1l_1+n_2l_2\cos\theta,x^2+n_2
l_2\sin\theta)$, with $n_1,n_2\in Z$, and contains the nontrivial
geometrical information concerning the angle of incidence. It is
the compactification of the plane where the interaction takes
place. The Dp-brane is then wrapping $T^{p-1}_{\|}$ and filling
the direction $x^1$ inside $T^2_\theta$, and these directions will
be decompactified at the end of the calculation. Note that in the
actual models of decaying mesons, the $x^2$ direction will be in
the transverse geometry and not in the Minkowski part of the
metric, which will be accounted for by $x^1$ and by some
dimensions of the (decompactification of the) $T^{p-1}_{\|}$
factor.
The macroscopic string is
winding along the other periodic direction on $T^2_\theta$, namely $x^2$, see figure \ref{torus}.

In the case of the string lying on a Dp-brane along a direction $X$, the compact 
space is $R_t \times X \times T^{p-1}_{\|} \times T^{9-p}_{\perp}$ where $X$ has the length of the string $L$.

In order to avoid unnecessary complications in writing down the
vertex operators, we adopt the trick \cite{jjp} of  choosing the
periodic direction along which the string is wrapped as a
``temperature'' direction, giving the opposite GSO projection of
the usual one, for which the ground states are scalars. Since we
are considering very long and therefore very massive strings, one
expects that the difference with respect to the usual GSO
projection is irrelevant, because it involves a finite number of
excitations only.

\begin{figure}[th]
\centerline{\psfig{file=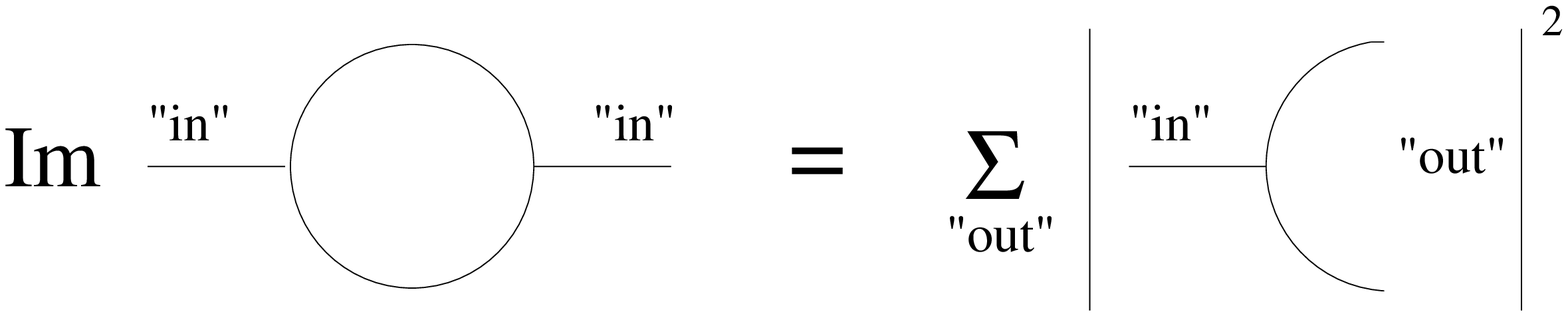,width=5.0in}} \vspace*{8pt}
\caption{The optical theorem: the imaginary part of the forward
amplitude is expressed as a the sum on the final states of the
decay. \protect\label{optical}}
\end{figure}

As to the final states after the splitting, one expects them to be
very excited, kinked strings, so their vertex operators are
presumably quite complicated. We are going to avoid the need of
writing down such operators by adopting an optical theorem, that
allows to sum over all the possible final states of the splitting,
giving the {\emph{total}} decay rate as the properly normalized
imaginary part of the `forward' amplitude, see figure
\ref{optical}.
To leading order in $g_s$, to which we limit ourselves, the total decay rate is just the one for the simple 
splitting, giving then the desired result.
The advantage of computing just the `forward' amplitude is that it involves the same simple vertex operators 
for the ``in'' states and the ``out'' states.
In our case it can be derived from the following correlator of two closed string vertex operators on the disk
\begin{equation}\label{corr}
{\cal A}=\langle {\cal V}_{(0,0)}(p_L,p_R){\cal V}_{(-1,-1)}(p_L^\prime,p_R^\prime)\rangle\ .
\end{equation}
Even if this is an open string process on the disk, the states are closed strings since the open strings wrap a 
compactified dimension.
From now on we will use a superscript ``$\theta$'' or ``$L$'' for quantities concerning the former and the latter case 
respectively, while formulas common to both cases will not have any superscript. The ground state vertex 
operators (for the tachyon) in the $(-1,-1)$ and $(0,0)$ pictures are given by \cite{jjp}
\begin{eqnarray}
&{\cal V}_{(-1,-1)}&=\frac{\kappa}{2\pi\sqrt{V}}:e^{-\phi-\tilde\phi+ip_L\cdot X+ip_R\cdot \tilde X}:\ ,\nonumber\\
&{\cal V}_{(0,0)}&=\frac{\kappa}{2\pi\sqrt{V}}\frac{\alpha^\prime}{2}(\psi\cdot p_L)(\tilde\psi\cdot p_R)
:e^{ip_L\cdot X+ip_R\cdot \tilde X}:\ .
\end{eqnarray}
The volume factor is $V^{\theta}=\sin\theta l_1 l_2 V_{\perp}V_{\|}$ for the case of the string intersecting the brane, and $V^L=L  V_{\|} V_{\perp}$ for the string lying on the brane, with $V_{\|}={\rm Vol}(T_{\|})$
and $V_\perp={\rm Vol}(T_{\perp})$.
It comes from the normalization of the amplitude with respect to the zero modes in the compact dimension.
In the formulas above, $\kappa$ is the gravitational constant (in the small energy limit the correlator (\ref{corr}) gives 
the propagation of the graviton), $\phi, \tilde\phi$ are the bosonized superghosts and $X, \tilde X, \psi,\ \tilde\psi$ the 
world-sheet bosons and fermions.
We need one fixed and one integrated vertex because the amplitude is on the disk.

The left and right momenta, on shell at the tachyon mass, have to
satisfy
 $p_L^2=p_R^2=\frac{2}{\alpha^\prime}$, $p_{L,R}=p\pm \vec L/2\pi\alpha^\prime$, with $\vec L^{\theta}=
 (0_t,l_2\cos\theta,l_2\sin\theta,\vec 0_{\|},\vec 0_{\perp})$ or $\vec L^{L}=(0_t,L,\vec 0_{\|}, \vec 0_{\perp})$.
In the case of the string intersecting the Dp-brane, we also allow
for a possible (almost continuous) velocity of the string along
$T_{\|}$, that is in the directions parallel to the brane and
orthogonal to the string world-sheet. The string momentum has then
the form
\begin{equation}
p^\theta=\frac{m}{\sqrt{1-v^2}}(1_t,0,0,\vec v_{\|},\vec 0)\ , \qquad {\rm{with}} \qquad m^2=\left(\frac{l_2}{2\pi\alpha^\prime}\right)^2-\frac{2}{\alpha^\prime} \ ,
\end{equation}
$m$ being the mass of the state, $\vec v \in T_{\|}$ and  $\vec 0\in T_{\perp}$.
At leading order in $g_s$, the Dp-brane is  a fixed background object and does not recoil, so in (\ref{corr}) one can 
take $|\vec v|=|\vec v^\prime|$.
Instead, for the string lying on the Dp-brane, we work in the rest frame of the string itself, so that
\begin{equation}
p^L=m(1_t,0,\vec 0_{\|}, \vec 0_{\perp})\ , \qquad\qquad m^2=\left(\frac{L}{2\pi\alpha^\prime}\right)^2-\frac{2}
{\alpha^\prime} \ .
\end{equation}

The  amplitude can be obtained by contracting all the fields in (\ref{corr}), using the usual formulas
\begin{eqnarray}
\langle X^\mu(z) X^\nu(z')\rangle = -\frac{\alpha'}{2}\eta^{\mu\nu}\log(z-z')\ , \quad
\langle X^\mu(z) \tilde X^\nu(\bar z')\rangle = -\frac{\alpha'}{2}G^{\mu\nu}\log(z-\bar z')
\end{eqnarray}
and the analogous ones for the fermions $\psi$ and the ghosts $\phi$.
The open string metrics read in the two cases
\begin{equation}
G^{\mu\nu,\theta} = {\rm diag}(-1_t,1,-1,I_{\|},-I_{\perp})\ , \qquad G^{\mu\nu,L} = {\rm diag}(-1_t,1,I_{\|},-I_
{\perp})\ .
\end{equation}
The invariants that appear in the amplitude are then
\begin{eqnarray}
-\sigma &\equiv &\frac{\alpha'}{2}p_L \cdot G \cdot p_R
=\frac{\alpha'}{2}p'_L \cdot G \cdot p'_R \ ,
\nonumber\\
-1-\frac{\alpha' t}{4}&\equiv & \frac{\alpha'}{2}p_L p'_L = \frac{\alpha'}{2}p_R p'_R\ ,
\nonumber\\
\sigma-\frac{\alpha' t}{4}&\equiv & \frac{\alpha'}{2}p_L \cdot G \cdot p'_R = \frac{\alpha'}{2}p'_L \cdot G 
\cdot p_R\ ,
\end{eqnarray}
(note that $p'_{L,R}=-p_{L,R}$) with $t=0,\ \sigma^\theta=-1+
\alpha' (l_2/2\pi\alpha')^2 \cos^2\theta,\ \sigma^L =
-1+\alpha^\prime (L/2\pi\alpha^\prime)^2$. The calculation
concerns macroscopically long strings, so the relevant limit is
that of large $l_2$ and $L$, hence large $\sigma^\theta \simeq
\alpha^\prime(l_2/2\pi\alpha^\prime)^2\cos^2\theta$  (unless
$\theta=\pi/2$) and large $\sigma^L \simeq
\alpha^\prime(L/2\pi\alpha^\prime)^2$. Although $t=0$, we keep
$t\neq 0$ as a regulator for the divergence in the real part of
the amplitude as $t\rightarrow 0$.

After the contractions are done, in order to obtain the amplitude ${\cal M}$ one faces the integral
\begin{equation}
\int_0^1 dx\, (1-x)^{-1-\alpha't/2}(1+x)^{1+2\sigma-\alpha't/2} x^{-1-\sigma}\ \sim \  2^{2\sigma} \frac
{\Gamma(-\alpha't/4) \Gamma(-\sigma)}{\Gamma(-\alpha't/4-\sigma)}\ ,
\end{equation}
where the approximate expression is valid as $t\rightarrow 0$.
The large $\sigma$ limit fluctuates wildly on the real axis, as one can see from the approximate expression,
since it contains the closed string state poles at integer values of $\sigma$ with zero width.
These fluctuations are averaged by taking the limit in a direction in the complex $\sigma$ plane at a small angle.
The infinitely narrow poles will then contribute with the proper weight to the imaginary part.
Practically, this amounts to applying Stirling's formula with the proper choice of phase, and results in
\begin{equation}\label{ampl}
{\cal M}\simeq -N_{D^2}\frac{\kappa^2}{(2\pi)^2 V}\frac{4(\sigma)^{1+\alpha^\prime t/4}}{\alpha^\prime t}
e^{-i\pi t\alpha^\prime/4}\ ,
\end{equation}
where  the normalization $N_{D^2}$ can be obtained by T-duality from the
standard partition function normalization $2\pi^2 V_9\tau_9$, giving $N_{D^2}^\theta=2\pi^2 l_1 V_{\|}\tau_p=
2\pi^2 l_1 V_{\|}/(2\pi)^{p} (\alpha^\prime)^{(p+1)/2}g_s$, $\ N_{D^2}^L=2 \pi^2L V_{\|}/(2\pi)^p(\alpha')^
\frac{p+1}{2} g_s$.

Using the optical theorem in order to extract the decay rate $\Gamma$ taking the imaginary part 
of (\ref{ampl}), $\Gamma=\frac{1}{m} {\rm Im} {\cal M}$, the singularity at $t=0$ is resolved.
Thus, the final results of the computation are
\begin{equation}\label{rate1}
\Gamma^{\theta} = \frac{g_s}{16\pi\sqrt{\alpha^\prime}}\cdot \frac{(2\pi\sqrt{\alpha^\prime})^{(8-p)}}
{V_\perp}\cdot\frac{\cos^2\theta}{\sin\theta}\ ,
\end{equation}
for the splitting rate of a string intersecting at an angle $\theta$ a generic Dp-brane,\footnote{Note that this quantity 
is finite and does not depend on the (transversal) velocity of the string, since it is computed in the 
rest frame of the latter.} and
\begin{equation}\label{rate2}
\Gamma^{L} = \frac{g_s}{32\pi^2\alpha^\prime}\cdot\frac{(2\pi\sqrt{\alpha^\prime})^{(9-p)}}{V_\perp}\cdot L\ ,
\end{equation}
for the splitting rate of a string of length $L$ lying on a generic Dp-brane.

The interpretation of the decay rates (\ref{rate1}), (\ref{rate2})
is the following. First of all, we have the natural
$(2\pi\sqrt{\alpha^\prime})^{(8-p)}/{V_\perp}$ or
$(2\pi\sqrt{\alpha^\prime})^{(9-p)}/{V_\perp}$ suppression given
by the transversal torus. It is due to the quantum delocalization
of the string in the directions transverse to the brane and it
just states that the distance between the string and the brane in
the transverse directions should be of order $\alpha'$ in order
for the interaction to take place. Second, in the rate
(\ref{rate1}) we have the factor $1/\sin\theta$ that describes the
fact that, when the string becomes more and more parallel to the
brane, the breaking probability increases, since the tension of
the string creates a bigger transversal force which helps the
string splitting. We also have the $\cos^2\theta$ term which is
the natural term symmetric as $\theta\rightarrow -\theta$ that
vanishes for the supersymmetric configuration  $\theta=\pi/2$, for
which the string does not split. In the rate (\ref{rate2}) we have
instead the $L$ factor, which is the phase space term, due to the
fact that since the string is entirely on the brane, it can split
at any point, so the rate is proportional to its length $L$.

Note that the calculation giving the rate (\ref{rate1}) is not
valid, strictly speaking, for the extreme values $\theta=0$ (by
construction: the torus used for the calculation becomes singular)
and $\theta=\pi/2$ (when we are no more in the Regge regime of
large $\sigma$). In the latter case the behavior of the resulting
rate is nevertheless the expected one. For $\theta\sim 0$,
instead, we observe that if we impose that the vanishing torus
direction, of length $L \sin\theta$, in the limit becomes one of
the transverse directions, we can write $V_{\perp(8-p)}=
V_{\perp(9-p)} / L \sin\theta$. By making this substitution in
(\ref{rate1}) we get the interpolating rate
\begin{equation}\label{rateinterpol}
\Gamma^{int} = \frac{g_s}{32\pi^2 \alpha'} \cdot \frac{(2\pi \sqrt{\alpha'})^{9-p}}{V_{\perp(9-p)}}\cdot L \cdot\cos^2\theta\ ,
\end{equation}
which for $\theta\rightarrow0$ exactly gives the rate (\ref{rate2}) as it should, since the string is ultimately lying on the Dp-brane for $\theta=0$.

\section{Holographic hadron decays}

In the previous section we have obtained very general results about the decay rate for the splitting of fundamental strings in flat space. We are now going to apply them in the study of some dynamical process involving mesonic states in the context of the gauge/string duality.

The basic idea is the following. In the gauge/gravity
correspondence one usually starts from a gauge theory engineered
using a stack of $N$ D-branes in some background. By taking the
large $N$ limit, the strong 't Hooft coupling regime of the gauge
theory is expected to be described by the near horizon limit of
the geometry created by the D-branes. Even if matter in the
fundamental representation can be  in principle incorporated in
this picture, this can be difficult to achieve in practice if the
number of flavors  is arbitrary. However, as discussed in
\cite{katz1}, when the number of flavors is small (with
respect to the number of colors $N$), we can study them from the holographic point
of view by adding appropriate probe branes in the supergravity
background dual to the theory without flavors. In this dual
description, mesons are described by open strings attached to the
flavor probe-branes. 
In particular one can have mesonic states of high
spin/energy which admit a semiclassical description as spinning
macroscopic fundamental strings. We are going to study decay
processes involving these kinds of states \footnote{Decays of small spin mesons have been 
discussed in  \cite{ss2,carlos}.}.

Let us discuss the general setting. The simplest supergravity
backgrounds dual to 4d field theories have metrics of the
form \bea\label{metric}
ds^2=e^{A(r)}(-dt^2+d\rho^2+\rho^2d\eta^2+dx_3^2)+e^{B(r)}dr^2+G_{ij}(r,\phi)d\phi^id\phi^j\
, \eea where $r$ and $\phi^i$, $i=1,\ldots,5$, describe
respectively the radial and the angular coordinates of  the six
dimensional internal space, and $r$ is associated (in a model
dependent way) to the energy scale of the dual theory whose UV and
IR regimes correspond to large and small $r$ respectively. Quite
generally, we picture the probe D-brane associated to the addition
of a flavor $Q$ as partly filling some internal angular
directions~$\chi^a$, while it is located at a fixed value of the
remaining angles $\psi^I_Q$. Furthermore, it fills the radial
coordinate from $r=\infty$ up to a point defined by  fixed angles
$\chi^a_Q$ and a minimal radius $r_Q$. This minimal radius is then
holographically associated to the flavor mass $m_Q$ in a model
dependent way. The fluctuations of the brane describe low mass
mesonic states, while mesonic states with very large spin can be
described by semiclassical spinning open strings with end-points
attached to the flavor D-brane.

We will now focus on the high spin mesons associated to rigid
spinning strings whose world-sheet is of the form \bea t=\tau\ ,\
\eta=\omega\tau\ ,\ r=r(\sigma)\ ,\ \rho=\rho(\sigma)\ ,\
\chi^a=\chi^a_Q\ ,\ \psi^I=\psi^I_Q\ . \eea The relevant equations
of motion can be easily derived from the effective action
\bea\label{effact} S=-\frac{1}{2\pi\alpha^\prime}\int d\tau
d\sigma
e^A(r)\sqrt{(1-\omega^2\rho^2)[(\rho^\prime)^2+e^{B(r)-A(r)}(r^\prime)^2]}\
, \eea supplemented by the boundary conditions
$r|_{\partial\Sigma}=r_Q$ and $\rho^\prime|_{\partial\Sigma}=0$
\cite{myers1}. For our purposes, it is convenient to  fix the
remaining reparameterization invariance by choosing the gauge
$\sigma=r$, so that the only effective dynamical field is
$\rho(\sigma)$. Then, if $r_0$ indicates the minimal radius
reached by the string, the energy and the spin of the meson are
given \cite{myers1} by
 $E=2F[r_0,r_Q]$ and $J=2H[r_0,r_Q]$ where
\bea
F[a,b]&=&\frac{1}{2\pi\alpha^\prime}\int_{a}^b d\sigma e^{A(\sigma)}\sqrt{\frac{(\rho^\prime)^2+
e^{B(\sigma)-A(\sigma)}}{1-\omega^2\rho^2}}\ ,\cr H[a,b]&=&\frac{\omega}{2\pi\alpha^\prime}\int_{a}^b d\sigma \rho^2
e^{A(\sigma)}\sqrt{\frac{(\rho^\prime)^2+ e^{B(\sigma)-A(\sigma)}}{1-\omega^2\rho^2}}\ .
\eea
One can  in principle invert these relations in order to get for example $E$ and $r_0$ as functions of the spin (and $r_Q$).

\begin{figure}[th]
\centerline{\psfig{file=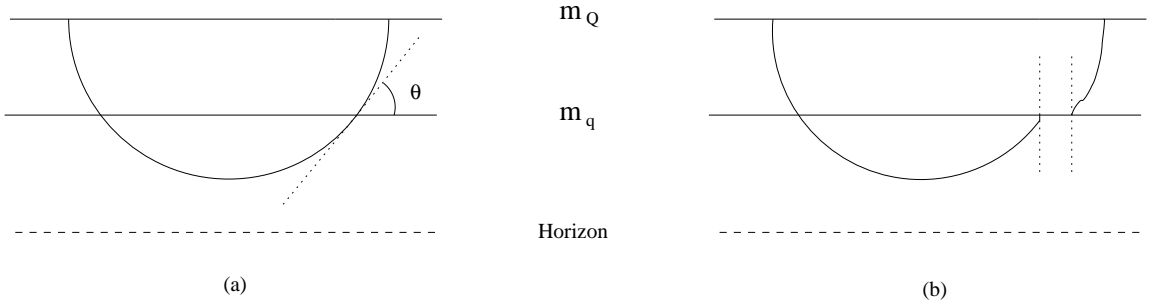,width=6.0in}} \vspace*{8pt}
\caption{(a) A large spin meson, bound state of two quarks of large mass $m_Q$, described by a string with both
end-points on the same brane. The string intersects a second brane, corresponding to lighter
quark masses $m_q$. (b) The strings after the splitting, representing two meson bound states of a heavy quark and a light quark. \protect\label{brdec2}}
\end{figure}

Let us consider now the effect of the introduction of another
D-brane associated to a lighter flavor $q$ of mass  $m_q<m_Q$,
i.e. with $r_q<r_Q$. We want to study the possible decay of the
above string, associated to a meson~$\bar Q Q$, into a couple of
strings representing the mesons $\bar Q q$ and $\bar q Q$, see figure \ref{brdec2}. 

We will
focus on decay rates that can be described within the
semiclassical picture, where the string classically intersects the
$q$-brane and then can split in the semiclassical regime. For
other kinds of mesons, whose D-brane is not aligned with the brane
corresponding to the heavier meson in such a way that the spinning
string intersects it, the meson decay involves world-sheet
instantonic transitions and then is exponentially suppressed in
the semiclassical regime. In order not to have this exponential
suppression we must then fulfill the conditions that $r_0[J]\leq
r_q$ and $\psi^I_q=\psi^I_Q$. If these conditions are satisfied,
our classical string can split into two open strings with
end-points attached to different branes which indeed correspond to
mesons of the kind ${\bar Q}q$ and ${\bar q}Q$. Rigid spinning
strings of this kind were studied in \cite{pt} and in our case we
expect our states to be some excited version of these rigidly
rotating strings, with also some linear momentum.

Even if we will not determine the explicit form of the outcoming
strings, it is important to note that their energies and total
angular momenta (computed with respect to the rest frame of the
initial meson) are completely determined by the classical
picture.  Indeed we can immediately
conclude that the lightest outcoming meson will have energy
$E_1=F[r_q,r_Q]$ and total angular momentum $J_1=H[r_q,r_Q]$,
while the heavier meson will have energy $E_2=E-E_1$ and angular
momentum $J_2=J-J_1$. The outcoming states will also have definite
and opposite linear momenta. If for example  $P^1$ and  $P^2$
denote the linear momenta of the lightest outcoming meson in the
directions \footnote{These are the coordinates in the metric
(\ref{metric}), not to be confused with the $x^{1,2}$ of the
previous section, that here correspond to $\rho$ and $r$
respectively.} $x_1=\rho\sin\omega\tau$ and
$x_2=\rho\cos\omega\tau$, we have that \bea P^1(t)&=&
\frac{\omega}{2\pi\alpha^\prime}\int_{r_q}^{r_Q} d\sigma
\rho\cos\omega\tau e^{A(\sigma)}\sqrt{\frac{(\rho^\prime)^2+
e^{B(\sigma)-A(\sigma)}}{1-\omega^2\rho^2}}\ ,
\eea
and $P^2$ can be obtained by the same expression by replacing  $\cos$ with $-\sin$.

Let us now see what we can say in general on the rate for such a
decay using the results obtained previously. Two basic ingredients
are the velocity $v$ of the string in the point where it splits
and its angle $\theta$ with the brane. These are given by \bea
v=\omega\rho(r_q)\quad,\quad
\cos^2\theta=\frac{(\rho^\prime(r_q))^2}{e^{B(r_q)-A(r_q)}+(\rho^\prime(r_q))^2}\
. \label{angol} \eea

In order to determine the decay rate, we have also to take into account the  suppression due to the effective transverse volume. This is a delicate point since 
such a transversal quantum delocalization of the string can be infinite \cite{susskind}.
In fact, if the string is free to sit at a generic point of a transverse dimension, quantum mechanically it is fully delocalized and the effective transverse volume in that direction is the whole length of the direction, which is infinite in the non compact case.
The situation is different if the string is classically at a fixed point of a direction, that is it sits at a minimum of a potential.
In this case the quantum delocalization, and so the effective length of the dimension, can be smaller.
The estimate of this effective size, which can be performed explicitly in the study of cosmic strings \cite{jjp}, is a non-trivial task in the present setting.

Finally, the decay rate is computed in flat space. When we go to (weakly) curved spaces, one has to replace $\alpha^\prime$ with an effective $\alpha^\prime_{eff}$ which depends on the warp factors of the metric.\\


\subsection{Meson decay in ${\cal N}=4$ Super Yang-Mills}
Of course, the equation of motion for $\rho(r)$ obtained from (\ref{effact}) is in general not analytically solvable and one must use some numerical or approximated method to evaluate it. We will now consider the most simple example where we can give an approximate analytical  estimate of the above observable quantities, namely the maximally supersymmetric case $AdS_5\times S^5$ with mesons of spin $J\gg \sqrt{\lambda}\ $ \cite{myers1}, where $\lambda=g_s N$ represents the 't Hooft coupling of the dual theory. In this case the flavor branes are D7-branes and it is convenient to use a different radial coordinate $z=R^2/r$, with $R^4=4\pi\alpha^{\prime 2}\lambda$ such that the relevant part of the metric is given by
\bea
ds^2=\frac{R^2}{z^2}(-dt^2+d\rho^2+\rho^2d\eta^2+dz^2)+\ldots\quad .
\eea
In this case the mass-radius relation is unambiguous and is given by $m_Q=R^2/2\pi\alpha^\prime z_Q$. As discussed in \cite{myers1}, in the case $J\gg\sqrt\lambda$, the spinning string solution is well approximated by a Wilson loop string \cite{malda} slowly spinning around its center of mass, i.e. $\rho(z)\simeq \rho_{st}(z)+\delta\rho(z)$ with very small $\delta\rho(z)$ and
\bea
\rho_{st}(z)=\int_z^{z_0}dx\frac{x^2}{\sqrt{z_0^4-x^4}}\ .
\eea
In this case  $\omega\ll 1$. Also, it is possible to show that in this limit
\bea
\omega^2\simeq  \frac{64{\cal C}^8m_Q^2}{\pi^2\lambda}\left(\frac{\lambda}{J^2}\right)^3 \quad,
\quad z_0^2\simeq \frac{\pi J^4}{16 {\cal C}^6m_Q^2\lambda}\ ,
\eea
 where ${\cal C}=\sqrt2 \pi^{3/2}/\Gamma(1/4)^2\simeq 0.599$.

If we now introduce a lighter  flavor by placing a second D7-brane (we will call the original brane ``Q-brane'' and the second one ``q-brane'')
at a position $z_q< z_0$, the spinning string does intersect it
and thus it can split. The condition that the decay is at all
possible, which in string theory terms is the fact that the
$q$-brane does intersect the string, $z_Q<z_q< z_0$, can be
expressed in terms of the particle properties as
\bea\label{mqcritical} 1> \frac{m_q}{m_Q}>\frac{4{\cal
C}^3\lambda}{\pi J^2}\ . \eea The existence of a limiting minimal
value of $m_q$ in order for the decay to happen can be understood
in field theory as follows. Since the theory is in a Coulomb
phase, the binding energy of the heavy quarks decreases as their
distance, and so their spin, increases. The total energy is
$E_{\bar Q Q} \sim 2m_Q -\alpha m_Q\lambda/J^2$, with $\alpha$
being some constant \cite{myers1}. On the other hand, the binding
energy $E^{bind}$ of the meson formed by a heavy and a light quark
in the limit of large quark separation is proportional to the mass
$m_q$ of the light quark and can be larger than $m_q$ in modulus
\cite{katz2,pt}. The total energy of the two mesons produced in
the decay would be $2E_{\bar Q q} \sim 2m_Q - E^{res}$ with
positive $E^{res}$. This is the strong coupling effect that makes
it possible for the heavy quark meson to decay, since for large
$J^2/m_Q\lambda$ it is possible that $E^{res}>\alpha
m_Q\lambda/J^2$, so that the total energy of the two produced
mesons is smaller than the one of the heavy quark meson
\cite{katz2}. This is the regime where the string does intersect
the $q$-brane in the dual setting. But, crucially, since $E^{res}$
is proportional to  $m_q$, for any fixed value of $m_Q\lambda/J^2$
there exists a minimal value of $m_q$ below which $E^{res}$ is
smaller than $\alpha m_Q\lambda/J^2$, forbidding the decay. This
critical value is precisely the one in (\ref{mqcritical}).

As we said before, in order for the decay not to be exponentially
suppressed, we must also require that the angular position of the
two branes in the transverse direction are equal, $\psi_q=\psi_Q$.
In the dual field theory, an unequal angle
$\Delta\psi=\psi_q-\psi_Q\neq 0$ would enter as a phase in the
coupling of one type of quark, let us say $q$,  with the complex
scalar of ${\cal N}=4$ SYM charged under $\psi$, schematically in
the superpotential as ${\cal W}=e^{i\Delta\psi}{\bar q}\Phi q$.
This phase suppresses the decay channel mediated by $\Phi$ and
ultimately should be responsible for the exponential suppression
of the decay rate. We do not venture at present to give  a precise
and explicit explanation of the suppression in field theory at
strong coupling.

Coming back to the string side of the duality, note that the velocity of the string at $z_q$ is of order $\sqrt{\lambda}/J$ and can be neglected in first approximation. Furthermore, if we restrict to the case in which the $q$ D7-brane is not ``too close'' to the $Q$ D7-brane (like for example if $z_Q/z_q=m_q/m_Q\sim \lambda/J^2$), then the decay rate is not completely suppressed since  the angle $\theta$ between the string and the brane is not too close to the value $\pi/2$ and can be evaluated to be
\bea
\theta\simeq{\rm arctg}\sqrt{\Big(\frac{\pi m_q J^2}{4 {\cal C}^3 m_Q\lambda}\Big)^4-1}\ .
\eea

The effective slope is given by $\alpha^\prime_{eff}=\frac{\pi^{-3/2}\sqrt{\lambda}}{2m_q^2}$.

Finally, in order to extract the total decay rate we need the
transversal volume $V_\perp$, that in this case is
one-dimensional. As we have already said, this is possibly the
most subtle point of the whole derivation. The string is
classically at a point of the transverse dimension, so its quantum
delocalization can be smaller than the size of the latter.
Contrary to what is done in \cite{jjp}, we cannot estimate
the delocalization with a local calculation around the
intersection point, since ultimately what generates the classical
localization are the boundary conditions on the $Q$-brane, which
fix the value of the angle $\psi_Q$ (locally, there is no
potential). So, since the complete calculation of the quantum
fluctuations around the classical string embedding seems
unfeasable at present, we will adopt a prudent choice that gives
as a natural (maximal) estimate a transversal length of order
$2\pi R\sim \sqrt{\alpha^\prime}\lambda^{1/4}$. We expect the
actual value of the delocalization to be of the same order. Then,
after taking into account the fact the the string can split at two
distinct points, we obtain the following minimal estimate of the
decay rate \bea \Gamma_{\bar Q Q\rightarrow \bar Q q+\bar q
Q}=\frac{m_q\sqrt{\lambda}}{8\sqrt{\pi}N\left(\frac{\pi
m_q}{4{\cal
C}^3m_Q}\right)^{2}\left(\frac{J^2}{\lambda}\right)^{2}\sqrt{\left(\frac{\pi
m_q}{4{\cal
C}^3m_Q}\right)^{4}\left(\frac{J^2}{\lambda}\right)^{4}-1}}\ .
\eea

The decay rate has precisely the expected behavior from the field theory point of view. It describes a $1/N$ process that increases as the coupling $\lambda$ increases. As the difference between the mass $m_q$ of the light quark and the mass $m_Q$ of the heavy quark becomes larger and larger, the decay is more and more probable.
However, there is a lower bound on this difference, below which the rate looses its meaning due to the square root. This lower bound is precisely the point at which the $q$ D7-brane reaches the lowest point of the string, below which there is no more intersection and therefore no decay.
Finally, the rate decreases as the spin $J$ of the heavy meson increases.
In fact,
increasing $J$ means increasing the distance between the two heavy quarks and since the theory is non confining, this reduces the binding energy and ultimately the energy density, making the decay process more and more disfavored.

\subsection{Meson decay in QCD-like theories}
Let us now try and put our general setting at work for models
which are a bit much closer to quenched QCD than the one discussed
above. Shortly after the $AdS/CFT$ correspondence was formulated,
a non singular gravity dual to a confining, non supersymmetric 4d
Yang-Mills theory was found in \cite{WYM}. The gauge
theory describes the low energy dynamics of a stack of $N\gg1$
D4-branes wrapped on a supersymmetry breaking circle and, in the
limit where the dual gravity description is reliable, it is
coupled with adjoint Kaluza-Klein fields. The background metric
and dilaton sourced by the D4-branes read
\bea
ds^2&=&(\frac uR)^{3/2} (dx_\mu dx^\mu + \frac{4R^3}{9u_h}f(u)d\theta_2^2) + (\frac{R}{u})^{3/2}  \frac{du^2}{f(u)} +
R^{3/2}u^{1/2} d\Omega_4^2\ ,\nonumber \\
e^{\Phi}&=&g_s\Bigl( \frac{u}{R}\Bigr)^{3/4}\ ,
\label{metricw}
 \eea
where $f(u)=(u^3-u_h^3)/u^3$. The radial coordinate $u$ is bounded from below by its ``horizon" value
$u_h$. 
String and field theory quantities are connected by the following relations \cite{bcmp,luca}:
$3u_h=\lambda m_0 \alpha',\ \lambda g^{-1}_s=3\pi N_c m_0 \sqrt{\alpha'},\
R^3=\lambda\alpha'/(3m_0),\
6\pi T=\lambda m_0^2$.
Here $\lambda=g^2_{YM}N_c$ is the 't Hooft coupling at the UV cut-off
and it has to be taken much greater than one in order for
the gravity approximation to be valid. Differently from pure Yang-Mills, the
theory has two different energy scales:
$T$, the Yang-Mills string tension, and $m_0$, the glueball and Kaluza-Klein mass scale.

The addition of $N_f\ll N$ flavors to the model above was realized in \cite{myers} by means of $N_f$
D6-brane probes and in \cite{ss} by means of $N_f$ D8 probes.
In the first model a generic D6 probe, embedded in the geometry
(\ref{metricw}), extends in the radial direction from a value
$u_Q$ up to infinity; the corresponding  quark has a constituent
mass which depends on $u_Q$ via the relation \be m_Q= {T\over
m_0}\int_{1}^{u_Q/u_h} dz \left[1-{1\over
z^3}\right]^{-{1\over2}}\ , \label{massq} \ee which is nothing but
the energy of an hypothetical string stretching from the horizon
at $u=u_h$ to $u= u_Q$. It turns out that in this model the
constituent quark mass cannot be zero, as the possible values
which $u_Q$ can take are bounded from below by a certain $u_{min}
> u_h$. The flavor symmetry in the model is $U(N_f)$ by
construction and the spontaneous breaking of the chiral $U(1)_A$
symmetry is accounted for by the bending of the flavor branes.

In the second model the D8-branes are curved, orthogonal to the circle $S^1(\theta_2)$ and extend up to the horizon
$u=u_h$; at large $u$ each curved D8-brane looks like a brane-antibrane pair. This picture provides a nice realization
of the dynamical UV restoration of the $U(N_f)\times U(N_f)$ chiral symmetry. The model describes
massless quarks.

We are now going to consider mesons with very high spin $J$ in the two models just
introduced. If $J\gg \lambda$ the strings associated to the mesons
can be studied semiclassically and the general expressions for their
splitting rates, as deduced above, can be used to extract the
corresponding exclusive meson decay rates. To study the physics of mesons built up by heavy
(light) quarks we will use the model with D6-brane (D8-brane) probes.

Heavy quarkonia with very high spin in the setup of \cite{myers} are described by macroscopic open strings,
with the extrema on a D6-brane at $u=u_Q$, which spin in the Minkowski directions and hang down with a $U$ shape up to a minimal
radial position $u=u_0$. The exclusive decay $Q\bar Q \rightarrow Q\bar q + q\bar Q$
is described by the splitting of the string on a second type of flavor D6 brane whose minimum is at a lower
position $u_q<u_Q$. The splitting can happen
only at one of the two intersection points and the decay rate
will be obtained by making use of our general formula (\ref{rate1}).
The decay is ``asymmetric'', in that the decay products are a  high spin ($J_1\gg\lambda$) meson
corresponding \cite{pt} to a string which bends down close to the horizon, and a meson with much smaller spin ($1\ll J_2\ll \lambda$),
corresponding to a string hanging down from one brane to the other
without approaching the horizon.

In the model of \cite{ss}, the open strings corresponding
to very light mesons extend and spin in the Minkowski directions,
thus lying on the flavor branes. This implies that there are
infinitely many points where the string can split, all along its
length. In this case the decay rate for a process like $q\bar
q\rightarrow q\bar q + \bar q q$ will be obtained by making use of
our formula (\ref{rate2}).

Let us start by estimating the decay rate for the process $Q\bar Q \rightarrow Q\bar q + q\bar Q$ in the
D6 model.
First of all we must put in eq. (\ref{rate1}) the corrected string tension and dilaton
to take care of the fact that we are not in flat space but on the
background (\ref{metricw}): $\alpha'\rightarrow \alpha'_{eff}=\alpha' (\frac{R}{u_q})^{3/2}$, and $g_s\rightarrow
e^{\Phi(u_{q})}=
g_s(\frac{u_q}{R})^{3/4}$.
Moreover we shall put $p=6$ and estimate the transverse volume as
\be
V_{\perp}=2\pi R_{\theta_2} \cdot 2\pi R_{S^4}= \frac{8\pi^2u_q}{3u_h^{1/2}}R^{3/2}f^{1/2}(u_q)\ .
\ee
In order to evaluate the $\theta$-dependent part in the decay rate, we need to know (see eq. (\ref{angol})
) the slope of the string profile at the intersection point. For this we do not have an analytic expression in general. However, for high spin mesons,
provided $u_Q\gg u_h$ and so (see formula (\ref{massq})) $m_Q \gg T/m_0$, it is possible \cite{crucco,pt} to approximate the
profile of the decaying open string with that corresponding
to a small perturbation
of a static, almost-$U$ shaped Wilson line \cite{pt}. Using the fact
that, in the semiclassical regime we work in ($J\gg\lambda$), the minimal radial distance $u=u_{0}$ reached by the ``$\bar Q Q$ string''
can be taken equal to $u_{h}$ up to exponentially suppressed terms \footnote{More
precisely \cite{sonne} $u_0\sim u_h[1+exp(-3m_0L/2)] $. In the
semiclassical regime the inter-quark distance $L$ , which increase with
$J$, is very large.} we can give the following expression for $r'(u)=r'_{st}(u)+\delta r'(u)$
\bea
r'_{st}(u) &=& (Ru_h)^{3/2}\frac{1}{(u^3-u_h^3)}\ , \nonumber \\
r'(u)&\approx& {(Ru_h)^{3/2}\over u_h^3(x^3-1)}\left[1-{x^3(x-1)\over y(x^3-1)}\right]\ , \quad x\equiv {u_q\over u_h}\ , \quad y\equiv {u_Q\over u_h}\ .
\eea

We have now all the data to put in our general formulas
(\ref{rate1}) and (\ref{angol}). The $\bar Q Q\rightarrow \bar Q q
+ \bar q Q$ decay rate of a large spin meson made up of heavy
quarks reads \be \label{ratef} \Gamma_{D6}={\lambda m_0\over
16\pi^2N}{\sqrt{x}\over(x^3-1)}\left[1+ {1\over
y}{(x-1)(1-2x^3)\over(x^3-1)}\right]\ . \ee It is possible to give
a clear interpretation of this formula, rewriting it in terms of
the constituent quark masses (\ref{massq}). Let us focus on two
special limits where we can have an analytic control of our
expressions.  The first amounts to taking
$u\gg u_h$ (large $x$), with $m_q\approx u_q/(2\pi\alpha')\gg T/m_0$. 
The resulting rate, expressed in terms of the quark masses, reads
\be
\label{finalrate1}
\Gamma_{D6} \sim \frac{\lambda}{16\pi^2 N} \left(\frac{T}{m_0}\right)^{5/2} \frac{m_0}{m_q^{5/2}}\left[1-2{m_q\over m_Q}\right]\ .
\ee
This expression depends on the two scales of the theory. In order to imagine how this could read in
a QCD-like theory, let us consider a limit where we identify the two scales taking $T\sim
m_0^2\sim\Lambda_{QCD}^2$;  this way
$
\Gamma_{D6} \sim\frac{\lambda}{N}\frac{\Lambda_{QCD}^{7/2}}{m_q^{5/2}}\left[1-2{m_q\over m_Q}\right].
$ We will refer to this formal limit as the ``QCD limit''.

The second limit on the masses amounts on taking $x\approx x_{min}$($\approx1.04$, see
\cite{crucco}).
This is the small mass limit where
\begin{equation}
m_q\approx \left({T\over m_0}\right){2\over\sqrt{3}}\sqrt{{u-u_h}\over u_h}\ .
\end{equation}
The decay rate now goes as
\begin{equation}
\label{finalrate2}
\Gamma_{D6} \sim {\lambda\over 36\pi^2N}\left({T\over m_0}\right)^2 {m_0\over m_q^2}\left[1-{T\over 3m_0 m_Q}\right]\ ,
\end{equation}
that in the ``QCD limit'' would read $\Gamma_{D6} \sim {\lambda\over N}
{\Lambda^3_{QCD}\over m_q^2}\left[1-{\Lambda_{QCD}\over m_Q}\right]$.

To get the decay rates in the rest frame of the laboratory, we
must multiply the obtained expressions by the relativistic time
dilation factor $\sqrt{1-v^2}$. As the decay can happen only
around the heavy quarks one can approximate with $L/2$ the
distance of the splitting point from the center of rotation, so
that $\sqrt{1-v^2}=\sqrt{1-(\omega L/2)^2}$. Then \cite{talk,noi2}
this factor reads $\sqrt{\frac{2m_Q/L}{T+2m_Q/L}}$, with $L$
proportional to some power of $J$.

Let us now comment on the decay rates we have found. They are
suppressed by $1/N$, grow with the coupling $\lambda$ and
increase as the mass of the produced quarks $m_q$ decreases: this
is indeed an expected behavior. Moreover the leading order
suppression of the rate with the mass $m_q$ is power-like, so the
one we have considered is the leading decay channel in the
QCD-like strongly coupled gauge theory at hand \footnote{Other
processes possibly involve instantonic world-sheet transitions and
are exponentially suppressed with $m_q$.}. The rates are mildly
dependent on the mass of the decaying mesons, which indirectly
enters in the formulas through the constituent quark mass $m_Q$.
The rates increase with this mass and go to a constant in the case
it is very large.
The spin $J$ enters in the expressions for the rates only through
the time dilation factor in the $2m_Q \ll LT$ regime, where it
suppresses the process. This is reminiscent of the suppression due
to the centrifugal barrier in some phenomenological models
\cite{spindep}. Instead, the corrections to the $u_{0}\sim u_{h}$
approximation are exponentially suppressed with $J$. As a final
remark, let us notice that the results we have obtained here for
heavy quarkonia apply to more general mesons made up of different
heavy quarks.

Let us now shift to the D8 model of \cite{ss} to study
light meson decays. In order to translate formula (\ref{rate2}) to
our case, note that \be \frac{g_s}{\alpha'} \rightarrow
\frac{e^{\Phi}}{\alpha'_{eff}}=\frac{g_s}{\alpha'}\left(
\frac{u_h}{R}\right)^{\frac{9}{4}}=
\frac{\lambda}{N}\frac{m_0^2\lambda^{3/2}}{3^{5/2}\pi} \ee and
that the strings are on the leading Regge trajectory \be
L=\sqrt{\frac{8 J}{\pi T}}=\frac{2 M}{\pi T}\ , \ee where $M$ is the
meson mass (the energy of the string). We then need to estimate
the suppression due to the transverse dimension. The procedure
proposed in \cite{jjp} consists on evaluating the quantum
delocalization of the string due to the quadratic fluctuations of
the world-sheet massive field associated to the transverse
direction. Taking the near horizon limit of the metric in
(\ref{metricw}) one can study the world sheet sigma model for the
transverse directions and discover that
their fluctuations create a broadening $\omega = \log [1+
(4R^{3/2}u_h^{1/2})/(9\alpha')]$.  In terms of field theory
quantities one can thus write \cite{noi2} \be \frac{(2\pi
\sqrt{\alpha'})^{9-p}}{V_{\perp}}=
\frac{2\pi}{\log^{1/2}(1+\frac{8\pi T}{9m_0^2})}\ . \ee We can now
put everything in  (\ref{rate2}), getting \be\label{gamma2}
\Gamma_{D8}= \frac{\lambda}{N}
\frac{1}{6\pi}\frac{1}{\log^{1/2}(1+\frac{8\pi
T}{9m_0^2})}\frac{T}{m_0}\sqrt{J}\ ,\ee or equivalently
\be\label{finalrate3} \Gamma_{D8}= \frac{\lambda}{N}
\frac{1}{6\pi \sqrt{2\pi}}\frac{1}{\log^{1/2}(1+\frac{8\pi
T}{9m_0^2})}\frac{\sqrt{T}}{m_0}M\ . \ee To get the rate per unit
length $L$ in the meson rest frame, we have to multiply the
expression above by the time dilation factor $\sqrt{1-v^2}$ and
then integrate along the length of the string. This only amounts
on multiplying the rate by a constant $\pi/4$ factor. In the ``QCD
limit" one just finds $\Gamma_{D8}\sim\lambda M/N$.

The result we have obtained for the rate has the expected scaling with $1/N$, and with the mass $M$ of the decaying meson.

\section{Discussion}
We have reviewed our attempts at coming to grips with
experimentally accessible predictions of string theory technology
as applied to models that view color tubes of QCD as strings, and
meson decays as a splitting of this string. It is encouraging that
rather specific results can be obtained, even if they are not yet
situated in completely realistic models. From the phenomenological
side, this field is dominated by the Lund
model \cite{Andersson:1998tv}, which has penetrated successfully
into widely used event generators. An important ingredient of this model, the Gaussian form of
the decay constant as a function of the mass of the quark pair
produced in the decay, does not seem to find an easy confirmation
in the theory. In this connection the attempt by \cite{zamaklar}
to provide such basis, in models that are similar to the ones used
in this paper, can be mentioned. There, it is linked to string
fluctuations, and in fact results from a Gaussian fit. We leave to
the future the task to resolve this issue, either by providing a
perhaps more profound string theory explanation, or possibly by
showing that phenomenology can be equally successful with a wider
range of functional dependencies.

\section*{Acknowledgments}
This work is partially  supported by the European Commission
contracts MRTN-CT-2004-005104, MRTN-CT-2004-503369, CYT FPA
2004-04582-C02-01, CIRIT GC 2001SGR-00065. F.~B., L.~M. and W.~T.
are supported by the Belgian Federal Office for Scientific,
Technical and Cultural Affairs through the ``Interuniversity
Attraction Poles Programme -- Belgian Science Policy" P5/27 (2006)
and P6/11-P (2007). F.~B. is in addition supported by the Belgian
Fonds de la Recherche Fondamentale Collective (grant 2.4655.07)
and the Belgian Institut Interuniversitaire des Sciences
Nucl\'eaires (grant 4.4505.86). L.~M. and W.~T. are also supported
by the FWO - Vlaanderen, project G.0235.05.

\end{document}